# Privacy Preservation by Disassociation

Manolis Terrovitis[*]
IMIS, Research Center 'Athena'
mter@imis.athena-innovation.gr

John Liagouris[†]
Dept. of Electrical and Comp. Eng., NTUA
liagos@dblab.ece.ntua.gr

Nikos Mamoulis
Dept. of Comp. Sci., Univ. of Hong Kong
nikos@cs.hku.hk

Spiros Skiadopoulos[‡]
Dept. of Comp. Sci., Univ. of Peloponnese
spiros@uop.gr

## ABSTRACT

In this work, we focus on protection against identity disclosure in the publication of sparse multidimensional data. Existing multi-dimensional anonymization techniques *(a)* protect the privacy of users either by altering the set of quasi-identifiers of the original data (e.g., by generalization or suppression) or by adding noise (e.g., using differential privacy) and/or *(b)* assume a clear distinction between sensitive and non-sensitive information and sever the possible linkage. In many real world applications the above techniques are not applicable. For instance, consider web search query logs. Suppressing or generalizing anonymization methods would remove the most valuable information in the dataset: the original query terms. Additionally, web search query logs contain millions of query terms which cannot be categorized as sensitive or non-sensitive since a term may be sensitive for a user and non-sensitive for another. Motivated by this observation, we propose an anonymization technique termed *disassociation* that preserves the original terms but hides the fact that two or more different terms appear in the same record. We protect the users' privacy by disassociating record terms that participate in identifying combinations. This way the adversary cannot associate with high probability a record with a rare combination of terms. To the best of our knowledge, our proposal is the first to employ such a technique to provide protection against *identity disclosure*. We propose an anonymization algorithm based on our approach and evaluate its performance on real and synthetic datasets, comparing it against other state-of-the-art methods based on generalization and differential privacy.

## 1. INTRODUCTION

The anonymization of sparse multidimensional data in the form of transactional data (e.g., supermarket sales logs, credit card logs, web search query logs) poses significant challenges. Adversaries that have a part of a record as background knowledge are aided by the dataset's sparsity in identifying the original record. Consider, for example, a dataset $D$ which contains records that trace web search query logs. Even without any direct identifier in the data (user's name or ID) the publication of $D$ might lead to privacy breaches, if an attacker has background knowledge that associates some queries to a known user. Assume that John knows that Jane was interested in buying air tickets to New York, so he has a background knowledge consisting of terms New York and air tickets. If $D$ is published without any modification then John can trace all records that contain both terms New York and air tickets. If only one such record exists, then John can easily infer that this record corresponds to Jane.

To counter such privacy leaks, several anonymization techniques have been proposed in the literature [5, 6, 11, 13, 14, 17, 19, 24, 27]. Most of these methods employ *generalization* [5, 13, 19, 27] to reduce the original term domain and eliminate identifying combinations. For example, they would generalize New York to North America, so that the infrequent combination would be replaced by the more frequent {North America, air tickets}. Alternatively, other methods which are based on suppression, simply remove infrequent terms or terms which participate in infrequent combinations. Generalization and suppression have been mostly used to provide protection against *identity* and *attribute* disclosure. There are few works that rely on adding noise (fake records or terms) to offer differential privacy [6, 14] or to hide the user intent in web search engines [24]. The problem with existing methods is that a large part of the the initial terms are usually missing from the anonymized dataset. There are only a few works [11, 18, 30] that preserve all original terms, without adding noise, based on the *Anatomy* [30] idea of separating quasi identifiers from sensitive values. Still, all these methods can only protect against attribute disclosure.

The main contribution of this work is a novel method called *disassociation* that preserves the original terms but hides identifying combinations. The privacy model is based on $k^m$-anonymity [27]: an adversary, who knows up to $m$ items from any record, will not be able to match his knowledge to less than $k$ records in the published data. Anonymization is achieved, not by hiding their constituent terms (as done by earlier approaches), but instead by suppressing the fact that some terms (like New York and air tickets) appear together in the same record. The *disassociation* transformation extends the idea of *Anatomy* [30] to provide for the first time protection against identity disclosure by separating terms of the original data. We focus on protection against *identity disclosure* for three reasons: *(a)* it is usually explicitly or implicitly required by law in many countries and applications, *(b)* it is often the case that the sensitivity of items cannot be accurately characterized, so protection against attribute disclosure is not an option, and *(c)* differential pri-vacy approaches [6, 14], which offer strong privacy protection, incur a high information loss that is often not an acceptable trade-off. Protecting identities using disassociation has already been identified as a complicated problem even for the case of simple relational data [18], and no previous solution exists for our problem settings. Finally, the proposed anonymization technique is equally capable to existing state-of-the-art methods in providing protection against attribute disclosure if sensitive terms have been identified.

---

[*]Supported by the EU/Greece funded COOPERATION: CAP Project.

[†]Supported by the EU/Greece funded Heracleitus II Program

[‡]Supported by the EU/Greece funded Thalis Program.

In brief, the main contribution of the paper is the proposal of *disassociation*, a new data transformation for sparse multidimensional data that preserves the original terms of the dataset. We show how this transformation can be used to anonymize a dataset and prove that the resulting data adhere to our privacy guarantee. Moreover, we present an anonymization algorithm that uses disassociation, and we show that it achieves limited information loss, by evaluating it experimentally on real and synthetic datasets.

## 2. PROBLEM DEFINITION

The proposed anonymization method focuses on sparse multidimensional data and provides protection against identity disclosure. This section formally presents our assumptions about the data and the attack model. In addition, we define the anonymity guarantee our method targets to. Figure 1 summarizes our notation.

**Data.** We assume a collection $D$ of *records*; each record is a set of *terms* from a huge *domain* $T$. For example, a term can be a query posed by a user in the context of web search logs, or a product bought by a customer in the context of supermarket logs. As a motivating example, consider a web search query log that traces the queries posed by users over a period of time. Each record models a different user and contains the set of queries posed by the user. Figure 2a presents an exemplary web search query log consisting of 10 records (each being the web search history of a different user). We do not distinguish between sensitive and non-sensitive terms; we consider the general case, where any term might reveal sensitive information for the user and any term can be used as part of a quasi-identifier for a user. As we discuss in Section 5, having a clear distinction between sensitive and identifying terms simplifies the problem and our proposed technique can guarantee, in this case, protection against attribute disclosure.

**Attack model.** The identification of a user in $D$ is made possible by tracing records that contain unique combinations of terms. For example, if the database of Figure 2a is published and an adversary $A$ knows that a user $U$ has searched for terms madonna and viagra, he can infer that only record $r_2$ contains both of them; therefore $A$ is certain that $r_2$ is associated to $U$. We assume that the adversary $A$ may have background knowledge of up to $m$ terms (i.e., queries) for any record (i.e., user) and that $A$ does not have negative background knowledge (i.e., the adversary does not know whether a user *did not* pose a specific query). Moreover, we assume that the adversary $A$ does not have background knowledge for so many individuals that it will allow her to infer negative knowledge about $U$ (see Section 5 for details).

**Anonymity guarantee.** The most popular guarantee for protection against identity disclosure is $k$-anonymity [26]. $k$-anonymity makes each published record indistinguishable from other $k-1$ published records. Applying $k$-anonymity on sparse multidimensional data can result to a huge information loss, since groups of identical records must be created in a sparse data space [1, 13, 28]. For this reason, we opt for $k^m$-anonymity [27], a conditional form of $k$-anonymity, which guarantees that an adversary, who has partial knowledge of a record (up to $m$ items, according to our assumption

| Symbol | Explanation | Symbol | Explanation |
|---|---|---|---|
| $D, D^A$ | Original, anonymized dataset | $T$ | Domain |
| $\mathcal{A}, \mathcal{I}$ | Anonymization, inverse transf. | $s(a)$ | Support of $a$ |
| $P / J \ldots$ | Clusters / Joint cluster | $T^P$ | cluster domain |
| $C, C_1, \ldots$ | Record Chunks | $T^C$ | Chunk domain |
| $SC, SC_1, \ldots$ | Shared chunks | $C_T$ | Term chunk |

**Figure 1: Notation**

above), will not be able to distinguish any record from other $k-1$ records. More formally:

DEFINITION 1. *An anonymized dataset $D^A$ is $k^m$-anonymous if no adversary that has a background knowledge of up to $m$ terms of a record can use these terms to identify less than $k$ candidate records in $D^A$.*

For the original dataset $D$ and its anonymized counterpart $D^A$, we define two transformations $\mathcal{A}$ and $\mathcal{I}$. The anonymization transformation $\mathcal{A}$ takes as input dataset $D$ and results in the anonymized dataset $D^A$. The inverse transformation $\mathcal{I}$ takes as input the anonymized dataset $D^A$ and outputs all possible (non-anonymized) datasets $D'$ that could produce $D^A$, i.e., $\mathcal{I}(D^A) = \{D' \mid D^A = \mathcal{A}(D')\}$. Obviously, $D \in \mathcal{I}(\mathcal{A}(D))$. For example, consider the dataset

$$D(age, zip) = \{(32, 14122), (39, 14122)\}$$

and its corresponding anonymized dataset (using generalization)

$$D^A(age, zip) = \{(3x, 14xxx), (3x, 14xxx)\},$$

we have: $\mathcal{A}(D) = D^A$ and

$$\mathcal{I}(D^A) = \left\{ \begin{array}{l} \{(30, 14000), (30, 14000)\}, \ldots \\ \{(30, 14000), (31, 14000)\}, \ldots \\ \{(32, 14122), (39, 14122)\}, \ldots \end{array} \right\}$$

In this paper, to achieve $k^m$-anonymity (Definition 1), we enforce the following privacy guarantee.

GUARANTEE 1. *Consider an anonymized dataset $D^A$ and a set $\mathcal{S}$ of up to $m$ terms. Applying $\mathcal{I}(D^A)$ will always produce at least one dataset $D' \in \mathcal{I}(D^A)$, for which there are at least $k$ records that contain all terms in $\mathcal{S}$.*

Intuitively, an adversary, who has limited background knowledge (consisting of a set $\mathcal{S}$ of up to $m$ terms) about a person, will have to consider $k$ distinct candidate records in a possible original dataset.

## 3. ANONYMIZATION BY DISASSOCIATION

In this paper, we propose an anonymization transformation $\mathcal{A}$ that is based on disassociation. The proposed transformation partitions the original records into smaller and disassociated subrecords. The objective is to hide infrequent term combinations in the original records by scattering terms in disassociated subrecords. To illustrate the crux of the disassociation idea, we will use Figure 2. We have already seen that the dataset of Figure 2a is prone to identity attacks (e.g., $r_2$ can be identified by madonna and viagra). The corresponding disassociated anonymized dataset is illustrated in Figure 2b. Our approach, initially, divides the table into two *clusters* $P_1$ and $P_2$ containing records $r_1$–$r_5$ and $r_6$–$r_{10}$ respectively. In each cluster $P_i$, records are partitioned to smaller subrecords, after dividing the terms in $P_i$ into subsets. For example, in $P_1$, the terms are divided into subsets $T_1 = \{$itunes, flu, madonna$\}$, $T_2 = \{$audi a4, sony tv$\}$, and $T_T = \{$ikea, viagra, ruby$\}$. Then, each record is split into subrecords according to these subsets. The collection of all subrecords of different records that correspond to the

|  | Record chunks |  | Term chunk |
| --- | --- | --- | --- |
|  | $C_1$ | $C_2$ | $C_T$ |
| $r_1$ | {itunes, flu, madonna} |  | ikea, viagra, ruby |
| $r_2$ | {madonna, flu} | {audi a4, sony tv} |  |
| $r_3$ | {itunes, madonna} | {audi a4, sony tv} |  |
| $r_4$ | {itunes, flu} |  |  |
| $r_5$ | {itunes, flu, madonna} | {audi a4, sony tv} |  |

Cluster $P_1$, $|P_1| = 5$

|  | Record chunk | Term chunk |
| --- | --- | --- |
|  | $C_1$ | $C_T$ |
| $r_6$ | {madonna, digital camera} | panic disorder, playboy, ikea, ruby |
| $r_7$ | {iphone sdk, madonna} |  |
| $r_8$ | {iphone sdk, digital camera, madonna} |  |
| $r_9$ | {iphone sdk, digital camera} |  |
| $r_{10}$ | {iphone sdk, digital camera, madonna} |  |

Cluster $P_2$, $|P_2| = 5$

| ID | Records |
| --- | --- |
| $r_1$ | {itunes, flu, madonna, ikea, ruby} |
| $r_2$ | {madonna, flu, viagra, ruby, audi a4, sony tv} |
| $r_3$ | {itunes, madonna, audi a4, ikea, sony tv} |
| $r_4$ | {itunes, flu, viagra} |
| $r_5$ | {itunes, flu, madonna, audi a4, sony tv} |
| $r_6$ | {madonna, digital camera, panic disorder, playboy} |
| $r_7$ | {iphone sdk, madonna, ikea, ruby} |
| $r_8$ | {iphone sdk, digital camera, madonna, playboy} |
| $r_9$ | {iphone sdk, digital camera, panic disorder} |
| $r_{10}$ | {iphone sdk, digital camera, madonna, ikea, ruby} |

(a) Original dataset $D$

(b) Anonymized dataset $D^A$

Figure 2: Disassociation example

same subset of terms is called a *chunk*. For example, $r_1$ is split into subrecords {itunes, flu, madonna}, which goes into chunk $C_1$ (corresponding to $T_1$), {}, which goes into chunk $C_2$, and {ikea, ruby}, which goes into chunk $C_T$. $C_T$ is a special, *term chunk*; the subrecords that fall into the last chunk ($C_T$) are merged to a single set of terms. In our example, $C_T$ contains set {ikea, viagra, ruby}, which represents the subrecords from all $r_1-r_5$ containing these terms. In addition, the order of the subrecords that fall in a chunk is randomized; i.e., the association between subrecords in different chunks is hidden. According to this representation, the original dataset could contain any record that could be *reconstructed* by a combination of subrecords from the different chunks plus *any subset of terms* from $C_T$. For example, {itunes, madonna, viagra, ruby} is a reconstructed record, which takes {itunes, madonna} from $C_1$, {} from $C_2$, and {viagra, ruby} from $C_T$. Observe that this record does not appear in the original dataset.

Similarly to the generalization based techniques, the disassociated dataset $D^A$ models a set of possible original datasets $\mathcal{I}(D^A)$. However, in our case the possible datasets are not described in a closed form captured by the generalization ranges, but by the possible combinations of subrecords. In other words, the original dataset is hidden amongst the multiple possible datasets in $\mathcal{I}(D^A)$ that can be reconstructed by combining the subrecords and terms taken from the disassociated dataset.

Overall, the anonymized dataset in Figure 2b satisfies Guarantee 1 for $k = 3$ and $m = 2$. We see in detail how this happens in Section 5, but we can observe that an attacker who knows up to $m = 2$ terms from a record $r$ of the original database is not able to reconstruct less that $k = 3$ records (by combining appropriate subrecords) that might have existed in the original data.

In the following, we present the details of our technique, which performs 3 steps: a horizontal partitioning, a vertical partitioning and a refining. The *horizontal partitioning* brings similar records together into clusters. The heart of the anonymization procedure lies in the *vertical partitioning* which disassociates infrequent combinations of terms. Finally, to reduce information loss and improve the quality of the anonymized dataset a *refining* step is executed.

**Horizontal partitioning.** Records of the original dataset $D$ are grouped into *clusters* according to the similarity of their contents (e.g., Jaccard similarity). For instance, cluster $P_1$ is formed by records $r_1-r_5$ (Figure 2b). Horizontal partitioning reduces the anonymization of the original dataset to the anonymization of multiple small and independent clusters. The benefits of this approach are threefold. First, it limits the scope of the term disassociation to the records that are contained in the cluster; two terms may be disassociated only within the local scope of a partition, limiting this way the negative effect in the information quality of the published dataset. Second, since clustering brings similar records together in the same partition, the anonymity guarantee can be achieved with reduced disassociation. Third, the anonymization process can be done more efficiently and even in parallel.

**Vertical partitioning.** Intuitively, vertical partitioning leaves term combinations that appear many times intact and disassociates terms that create infrequent and, thus, identifying combinations. The disassociation is achieved by concealing the fact that these terms appear together in a single record. Vertical partitioning applies on each cluster and divides it into *chunks*. There are two types of chunks: record and term chunks. *Record chunks* contain subrecords of the original dataset; i.e., each chunk is a collection (with bag semantics) of sets of terms, and they are $k^m$-anonymous. That is, every $m$-sized combination of terms that appears in a chunk, appears at least $k$ times. *Term chunks* do not contain subrecords; they contain the terms that appear in the cluster, but have not been placed to record chunks. A term chunk is a simple collection of terms with set semantics. Each cluster may contain an arbitrary number of record chunks ($\geq 0$) and exactly one term chunk (which might be empty). In Section 5, we explain how the term chunk can be used to provide $l$-diversity some terms have been designated as sensitive.

Vertical partitioning is applied to each cluster independently. Let us consider a cluster $P$ and let $T^P$ be the set of terms that appear in $P$. To partition $P$ into $v$ record chunks $C_1, \ldots, C_v$ and a term chunk $C_T$, we divide $T^P$ into $v+1$ subsets $T_1, \ldots, T_v, T_T$ that are pairwise disjoint (i.e., $T_i \cap T_j = \emptyset$, $i \neq j$) and jointly exhaustive (i.e., $\bigcup T_i = T^P$). Subsets $T_1, \ldots, T_v$ are used to define record chunks $C_1, \ldots, C_v$ while subset $T_T$, is used to define term chunk $C_T$. Specifically, $C_T = T_T$ and record chunks $C_i$, $1 \leq i \leq v$ are defined as $C_i = \{\!\{ T_i \cap r \mid \text{for every record } r \in P \}\!\}$ where $\{\!\{\cdot\}\!\}$ denotes a collection with bag semantics (i.e., duplicate records are allowed in $C_i$). Thus, chunks $C_1, \ldots, C_v$ are collections of records while chunk $C_T$ is a set of terms. The partitioning of $T^P$ to $T_1, \ldots, T_v, T_T$ is performed in a way which ensures that all resulting record chunks $C_1, \ldots, C_v$ are $k^m$-anonymous. In Figure 2b, two $3^2$-anonymous record chunks $C_1$ and $C_2$ are formed for $P_1$, by projecting the records of $P_1$ to sets $T_1 = \{$itunes,flu,madonna$\}$ and $T_2 = \{$audi a4, sony tv$\}$ respectively; the remaining terms {ikea, viagra,ruby} of $P_1$ form the term cluster $C_T$.

Note that, for each published cluster, we explicitly show the number of original records in it. Without this explicit information, a data analyst may only infer that the cluster has at least as many records as the cardinality of the chunk with the greatest number of subrecords. Not knowing the cardinality of a cluster introduces significant information loss; for instance, it is not feasible to estimate the co-existence of terms in different chunks.

Finally, we remark that horizontal and vertical partitioning are applied in reverse order from what is followed by approaches that employ similar data transformations [11, 18, 30]. Thus, since verti-

946

| Record | | Term | Shared |
|---|---|---|---|
| $P_1$ cluster | | | |
| {itunes, flu, madonna} | | | {ikea,ruby} |
| {madonna, flu} | {audi a4, sony tv} | viagra | {ruby} |
| {itunes, madonna} | {audi a4, sony tv} | | {ikea} |
| {itunes, flu} | | | |
| {itunes, flu, madonna} | {audi a4, sony tv} | | |
| $P_2$ cluster | | | |
| {madonna, digital camera} | | | {ikea,ruby} |
| {iphone sdk, madonna} | | panic | |
| {iphone sdk, digital camera, madonna} | | disorder, | |
| {iphone sdk, digital camera} | | playboy | |
| {iphone sdk, digital camera, madonna} | | | {ikea,ruby} |

**Figure 3: Disassociation with a shared chunk.**

cal partitioning is applied *independently* in each horizontal partition (i.e., cluster), our method follows a *local* anonymization approach. This constitutes a significant difference from previous works that anonymize datasets by performing a global partitioning between terms (usually between sensitive terms and quasi-identifiers).

**Refining.** At this final step of the method, we focus on improving the quality of the published result while maintaining the anonymization guarantee. To this end, we examine the terms that reside in term chunks. Consider the example of Figure 2b. Terms ikea and ruby are in the term chunk of $P_1$ because their support in $P_1$ is low (each term appears in only 2 records). For similar reasons these terms are also in the term chunk of $P_2$. However, the support of these terms considering both clusters $P_1$ and $P_2$ is not small enough to endanger user privacy (ikea and ruby appear as many times as itunes and iphone that are in record chunks).

To address such situations, we introduce the notion of *joint clusters* that offer greater flexibility in our partitioning scheme by allowing different clusters to *share* record chunks. Given a set $T^s$ of *refining terms* (e.g., ikea and ruby), which commonly appear in the term chunks of two or more clusters (e.g., $P_1$ and $P_2$), we can define a joint cluster by *(a)* constructing one or more *shared chunks* after projecting the original records of the initial clusters to $T^s$ and *(b)* removing all $T^s$ terms from the term chunks of the initial clusters. Figure 3 shows a joint cluster, created by combining clusters $P_1$ and $P_2$ of Figure 2b, based on $T^s$={ikea,ruby}.

The idea of a joint cluster can be recursively generalized. We may form higher-level joint clusters by combining simple and joint clusters of a lower level (for example see Figure 5). In the general case a joint cluster $J$, has as children the joint clusters $J_1, \ldots, J_n$ and at its leaves the simple clusters $P_1, \ldots, P_m$. Moreover it contains the $k^m$-anonymous shared chunks $SC_1, \ldots, SC_w$, which are created over a domain $T^s$. All terms of $T^s$ come from the term chunks $C_{T_1}, \ldots, C_{T_m}$ of $P_1, \ldots, P_m$. If $T_1, \ldots, T_w$ are the domains of $SC_1, \ldots, SC_w$, $T_1 \cup \cdots \cup T_w = T^s$ and $T_i \cap T_j = \emptyset$ for $i \neq j$, then each shared chunk $SC_i$ is created by projecting the records of every $P_j$ to $C_{T_j} \cap T_i$. Shared chunks are defined in this way, in order to avoid having a record contributing the same projection to shared or simple record chunks more than once.

**Reconstruction of datasets.** A disassociated dataset $D^A$ has the original records of $D$ partitioned into subrecords (residing in record or shared chunks) and terms (residing in term chunks). An adversary $A$ can combine record, shared and term chunks in an effort to reconstruct the world of all possible original datasets $\mathcal{I}(D^A)$. Possible original datasets may be reconstructed by combining the subrecords of record and term clusters padded with some terms from the term chunks. Such datasets $D'$ are called *reconstructed datasets* and by construction belong to $\mathcal{I}(D^A)$. The adversary $A$ may consider only the reconstructed datasets that abide to his background knowledge. Guarantee 1 requires that for every $m$ terms that exist in a record of $D$, there will be a $D'$ that contains $k$ records with these terms. Thus, an adversary will always have $k$ candidate records that will match her background knowledge.

Reconstructed datasets are also useful to data analysts, since they have similar statistical properties to the original one. We experimentally evaluate this similarity in Section 7. The benefit of providing the disassociated form, instead of a reconstruction directly, is threefold: *(a)* an analyst can work directly on the disassociated dataset. The disassociated dataset reveals some information, i.e., itemset supports, that is *certain* to exist on the original data, *(b)* the reconstruction procedure is transparent; an adversary cannot draw incorrect conclusions about the identity of a user by considering the reconstructed dataset as original or as ineffectively perturbated and *(c)* an analyst can create an arbitrary set of reconstructed datasets and average query results from all of them.

## 4. THE ANONYMIZATION ALGORITHM

The proposed algorithm uses heuristics to perform the partitioning (horizontal and vertical) and the refining step of Section 3.

**Horizontal partitioning.** Horizontal partitioning should bring together similar records that contain many common terms and few uncommon ones. Similarity may be assessed using measures from Information Theory (e.g., Jaccard coefficient). Related clustering algorithms exist in the literature for set-valued data [29], but unfortunately they are not appropriate for our setting since: *(a)* they are not efficient on large datasets and *(b)* they do not explicit control the size of the clusters. We employ Algorithm HORPART, a lightweight heuristic that does not have these problems. The key idea is to split the dataset into two parts: one with the records that contain the most frequent term a in the dataset and another with the remaining records. This procedure is recursively applied to the new datasets until the final datasets are small enough to become clusters. Terms that have been previously used for partitioning are recorded in set *ignore* and are not used in subsequent splitting (Line 3).

**Vertical partitioning.** To vertically partition the clusters, we follow a *greedy* strategy (Algorithm VERPART), executed independently for each cluster. VERPART takes as input a cluster $P$ and integers $k$ and $m$; the result is a set of $k^m$-anonymous record chunks $C_1, \ldots, C_v$ and the term chunk $C_T$ of $P$.

Let $T^P$ be the set of terms of $P$. Initially, the algorithm computes the number of appearances (support) $s(t)$ of every term $t$ and orders $T^P$ with decreasing $s(t)$. All terms that appear less than $k$ times are moved from $T^P$ to the term chunk $T_T$. Since all the remaining terms have support at least $k$, they will participate in some record chunk. Next, the algorithm computes sets $T_1, \ldots, T_v$ (while loop). To this end, the algorithm uses set $T_{remain}$ that contains the non-assigned terms (ordered by decreasing support $s$) and $T_{cur}$ (that contains the terms that will be assigned to the current set). To compute $T_i$ ($1 \leq i \leq v$), Algorithm VERPART considers all terms of set $T_{remain}$. A term $t$ is inserted into $T_{cur}$ only if the $C_{test}$ chunk constructed from $T_{cur} \cup \{t\}$ remains $k^m$-anonymous (Line 12). Note that the first execution of the for loop (Line 10) will always add a term $t$ to $T_{cur}$ since $C_{test} = \{t\}$ is $k^m$-anonymous

**Algorithm**: HORPART
**Input** : Dataset $D$, set of terms *ignore* (initially empty)
**Output** : A HORizontal PARTitioning of $D$
**Param.** : The maximum cluster size $maxClusterSize$

1 **if** $|D| < maxClusterSize$ **then return** $\{\{D\}\}$;
2 Let $T$ be the set of terms of $D$;
3 Find the most frequent term a in $T - ignore$;
4 $D_1 =$ all records of $D$ having term a;
5 $D_2 = D - D_1$;
6 **return** HORPART($D_1$, *ignore* $\cup$ a)$\cup$HORPART($D_2$, *ignore*)

```
Algorithm: VERPART
Input   : A cluster P, integers k and m
Output  : A k^m-anonymous VERtical PARTitioning of P
```
1  Let $T^P$ be the set of terms of $P$;
2  **for** *every term* $t \in T^P$ **do**
3      Compute the number of appearances $s(t)$;
4  Sort $T^P$ with decreasing $s(t)$;
5  Move all terms with $s(t) < k$ into $T_T$;        //$T_T$ is finalized
6  $i = 0$;
7  $T_{remain} = T^P - T_T$;          //$T_{remain}$ has the ordering of $T^P$
8  **while** $T_{remain} \neq \emptyset$ **do**
9      $T_{cur} = \emptyset$;
10     **for** *every term* $t \in T_{remain}$ **do**
11        Create a chunk $C_{test}$ by projecting to $T_{cur} \cup \{t\}$ ;
12        **if** $C_{test}$ *is* $k^m$-*anonymous* **then** $T_{cur} = T_{cur} \cup \{t\}$;
13     $i$++;
14     $T_i = T_{cur}$;
15     $T_{remain} = T_{remain} - T_{cur}$;
16 Create record chunks $C_1, \ldots, C_v$ by projecting to $T_1, \ldots, T_v$;
17 Create term chunk $C_T$ using $T_T$;
18 **return** $C_1, \ldots, C_v, C_T$

```
Algorithm: REFINE
Input   : A set P of k^m-anonymous clusters
Output  : A refinement of P
```
1  **repeat**
2      Add to every joint cluster a virtual term chunk as the union of the term chunks of its simple clusters;
3      Order (joint) clusters in $\mathcal{P}$ according to the contents of their (virtual) term chunks;
4      Modify $\mathcal{P}$ by joining adjacent pairs of clusters (simple or joint) based on Equation 1;
5  **until** *there are no modifications in* $\mathcal{P}$;
6  **return** $\mathcal{P}$

($s(t) \geq k$). If the insertion of a term $t$ to $T_{cur}$ renders $T_{cur} \cup \{t\}$ non $k^m$-anonymous, $t$ is skipped and the algorithm continues with the next term. After having assigned to $T_{cur}$ as many terms from $T_{remain}$ as possible, the algorithm *(a)* assigns $T_{cur}$ to $T_i$, *(b)* removes the terms of $T_{cur}$ from $T_{remain}$ and *(c)* continues to the next set $T_{i+1}$. Finally, Algorithm VERPART constructs record chunks $C_1, \ldots, C_v$ using $T_1, \ldots, T_v$ and the term chunk $C_T$ using $T_T$.

**Refining**. The result of the vertical partitioning is a set $\mathcal{P}$ of $k^m$-anonymous clusters. The refining step improves the quality of the anonymized dataset by iteratively creating joint clusters until no further improvement is possible. A naïve method to perform this step consists of computing the information loss (e.g., using a metric of Section 6) for all possible refinement scenarios and choosing the one with the best effect on data quality. Since such an option is very inefficient, we define a refining criterion. Let us consider two clusters $J_1$ and $J_2$. These cluster are joined into cluster $J_{new}$ if the following inequality holds:

$$\frac{s(t_1) + \cdots + s(t_n)}{|J_{new}|} \geq \frac{u_1 + \cdots + u_m}{|P_1| + \cdots + |P_m|} \quad (1)$$

where *(a)* $t_1, \ldots, t_n$ are the refining terms $T^s$ (Section 3), *(b)* $s(t_1), \ldots, s(t_n)$ are the supports of $t_1, \ldots, t_n$ respectively in the shared chunks of $J_{new}$, *(c)* $P_1, \ldots, P_m$ are the simple clusters of $J_1$ and $J_2$ that contain $t_1, \ldots, t_n$ and *(d)* $v_1, \ldots, v_m$ are the number of terms $t_1, \ldots, t_n$ that appear in the term chunk of each of $P_1, \ldots, P_m$ respectively. For instance, if $J_1$ and $J_2$ are clusters $P_1$ and $P_2$ of Figure 2b and $J_{new}$ is the joint cluster of Figure 3 then the refining terms are ruby and ikea and we have: $\frac{s(\text{ruby}) + s(\text{ikea})}{|J_{new}|} = \frac{4+4}{10} \geq \frac{2+2}{10} = \frac{u_1 + u_2}{|P_1| + |P_2|}$. Thus, $J_1$ and $J_2$ are replaced by $J_{new}$.

Note that the left part of Equation 1 estimates the probability of attributing one of $t_1, \ldots, t_n$ to the records of the joint $J_{new}$ while the its right part expresses the probability of attributing one of $t_1, \ldots, t_n$ to the initial records of $J_1$ and $J_2$.

Even with the criterion of Equation 1, we still need to exhaustively explore all the combinations of clusters (simple or joint) in order to choose the best ones. This is computationally infeasible. Thus, we have opted for a heuristic that merges each time only two existing clusters (simple or joint) to form a new joint cluster. The sketch of this method is illustrated in Algorithm REFINE. The algorithm takes as input a collection of simple clusters $\mathcal{P}$ and transforms it to a collection of joint clusters. The algorithm orders the clusters of $\mathcal{P}$ as follows: a) each term t is given a *term chunks support* $tcs(\text{t})$; i.e., the number of term chunks in clusters of $\mathcal{P}$ where t appears; b) the terms in term chunks are ordered in descending order of their $tcs$; and c) clusters are ordered by comparing lexicographically their term chunks. After the first iteration, joint clusters are introduced in $\mathcal{P}$. To each joint cluster $J$, we add a *virtual term chunk*, which is the union of the term chunks of its simple clusters, and we use it in the ordering step. REFINE modifies $\mathcal{P}$ by merging adjacent pairs of clusters and repeats the process until $\mathcal{P}$ does not change. The merging is done only if the criterion of Equation 1 is satisfied, and produces a joint cluster as defined in Section 3.

**Correctness of the algorithm.** Disassociation performs the partitioning (vertical and horizontal) and refining steps detailed in the previous sections. The proposed method is correct; it succeeds for any input and it always produces a disassociated $k^m$-anonymous dataset. It is not hard to verify that the algorithm terminates and produces a disassociated result. The proof that a disassociated dataset is $k^m$-anonymous is provided in Section 5. In a nutshell notice that *(a)* horizontal partitioning does not alter the original dataset and always produces clusters, *(b)* vertical partitioning creates $k^m$-anonymous clusters since Algorithm VERPART will put every term that has support over $k$ to the record chunks (Lines 10-17) and the rest of the terms in the term chunk (Lines 6 and 18) and *(c)* refining has the trivial solution of not merging any clusters and if a joint cluster is created (i.e., if shared chunks are added), then $k^m$-anonymity is preserved as we prove in Section 5.

**Complexity.** The most expensive part of disassociation is the horizontal partitioning that has a worst case complexity of $O(|D|^2)$ time. The horizontal partitioning can be seen as a version of quicksort, which instead of a pivot uses the most frequent term to split each partition; in the worst case it will do $|D|$ partitionings and at each partitioning it has to re-order $|D|$ records. The complexity of vertical partitioning depends on the domain $T^P$ of the input cluster $P$, and not on the characteristics of the complete dataset. The most expensive operation in the vertical partitioning is to establish whether a clunk is $k^m$-anonymous or not. This task requires examining $\binom{|T^P|}{m}$ combinations, thus it takes $O(|T^P|!)$ time. Since we regulate the size of clusters, the behavior of the overall algorithm, as the dataset size grows, is dominated by the behavior of the horizontal partitioning. Finally, the refining algorithm has again a $O(|D|^2)$ time complexity, since in the worst case it will perform as many passes over the clusters as the number of the clusters. Note that this a worst case analysis; in practice, the behavior of our algorithm is significantly better; this is also verified by the experimental evaluation of Section 7, which shows a linear increase of the computational cost with the input dataset size $|D|$.

## 5. ANONYMIZATION PROPERTIES

In Section 3, we described our disassociation transformation technique, which is implemented by the algorithm presented in Section

| Records | Record chunks | | Term chunk |
|---|---|---|---|
| | $C_1$ | $C_2$ | $C_T$ |
| a | a | | |
| a | a | | |
| b, c | a | b, c | |
| b, c | | b, c | |
| a, b, c | | b, c | |
| (a) Original dataset | (b) Anonymized dataset | | |

**Figure 4: Illustration of Example 1, Original cluster size = 5**

4. In this section, we prove how the disassociated result can guarantee $k^m$-anonymity, by showing how the transformed data can be used to reconstruct a possible initial dataset that contains $k$ times any combination of $m$ terms. In this proof we define two additional properties that must be preserved in a disassociated dataset.

**Cluster anonymity.** First, we prove that each disassociated cluster is $k^m$-anonymous, by constructing an initial cluster that contains $k$ times any $m$ terms of the disassociated cluster.

Let $P$ be an arbitrary cluster of the anonymized dataset which is vertically partitioned into $k^m$ anonymous record chunks $C_1, \ldots, C_v$ and a term chunk $C_T$. Then the following Lemma holds:

LEMMA 1. *For any $m$ terms $\mathcal{S} = t_1, \ldots, t_m$ that appear in $P$, at least $k$ distinct records that contain $\mathcal{S}$ can be reconstructed by combining subrecords from the chunks $C_1, \ldots, C_v$ and terms from $C_T$, or no record can be reconstructed that contains $\mathcal{S}$.*

PROOF. We first prove that Lemma 1 holds if all $m$ terms fall inside the record chunks. In this case the $m$ terms $\mathcal{S} = t_1, \ldots, t_m$ are scattered in $n, (n \leq m, n \leq v)$ record chunks $C_1, \ldots, C_n$. Let $S_1, \ldots, S_n$ be the subsets of $\mathcal{S}$ that appear in each of $C_1, \ldots, C_n$. Since each chunk is $k^m$ anonymous, $S_i$ will appear in the respective record chunk $C_i$ at least $k$ times together or none at all. The latter case happens if the $S_i$ terms exist in disjoint groups of subrecords inside $C_i$. If there is even one of $S_1, \ldots, S_n$ whose terms do not appear together at all in the respective chunk, then the $\mathcal{S}$ terms cannot appear together in any reconstructed record. If every set of $S_1, \ldots, S_n$ appears together in the respective chunk, then it has to appear in at least $k$ subrecords in each chunk. Let $SR_1, \ldots, SR_n$ be these sets of subrecords, one from each record chunk. We can then create a record by combining 1 subrecord from each of $SR_1, \ldots, SR_n$, i.e., $r = sr_1 \cup \cdots \cup sr_n$, where $sr_i$ is a subrecord, $sr_i \in SR_i$. Since each $SR_i$ contains at least $k$ subrecords, we remove the used subrecord and repeat the process at least $k-1$ more times. This results to at least $k$ distinct records that contain all $\mathcal{S}$ terms. Assume now that only $g$, $g < m$ terms fall inside the record chunks and $m - g$ terms fall in the term chunk. The previous proof holds for the $g$ terms too, since $g < m$, thus $k$ records can be reconstructed from the record chunks that contain the $g$ items. We can then directly pad these $k$ records with the rest of $m - g$ terms from the term chunk. We are free to do so, since the multiplicity and the correlations of these terms are not disclosed in the disassociated cluster. □

Lemma 1 shows that $k$ records can be constructed from a disassociated cluster; still, this is not sufficient for providing $k^m$-anonymity as defined in Guarantee 1 as the following example illustrates.

EXAMPLE 1. *Let us consider the dataset of Figure 4a. Assume that we want to publish it as $3^2$-anonymous and that we create two record chunks $C_1$ and $C_2$ with domains $T_1 = \{a\}$, $T_2 = \{b,c\}$ and $T_T = \emptyset$, as illustrated in Figure 4b. It is not hard to verify that all chunks are $3^2$-anonymous and that Lemma 1 holds.*

*Let us now consider an adversary A that knows: (a) the anonymized dataset of Figure 4b, (b) that the size of the original cluster is 5 and (c) that a user had used terms a and b, i.e., $\{a,b\}$ is a subrecord of the original dataset. Adversary A also knows that the original dataset is composed by a combination of the records stored in chunks $C_1$ and $C_2$.*

*While the subrecords from $C_1$ and $C_2$ can be combined to create $k = 3$ records that contain a and b, these records cannot appear in any original dataset, which must contain 5 records. It is not hard to verify that the only combination that results in a dataset with 5 records is the one presented in Figure 4a. Thus, no dataset that contains $\{a,b\}$ 3 times can be the initial dataset of the example of Figure 4a. This way, the user's record $\{a,b,c\}$ is revealed.*

Example 1 demonstrates that Lemma 1 is not sufficient to guarantee $k^m$-anonymity. Lemma 1 guarantees that $k$ records that contain any $m$ terms can be constructed, but it does not guarantee that these records can appear in a valid dataset of a predefined size. The sparsity of the original data, often leads to empty subrecords inside different chunks. Since there cannot be empty records, a record that is created as a result of combining empty subrecords is not valid. An initial dataset that contains an empty record is also not valid, thus and adversary can discard it. To enforce Guarantee 1, we must require not only that Lemma 1 holds, but also that these records can appear in a valid initial dataset. Fortunately, we do no need to reconstruct all possible original datasets to see if this condition is satisfied. It suffices to enforce the condition of the following lemma.

LEMMA 2. *Let $C_1, \ldots, C_v$ be the record chunks that correspond to the anonymization of a cluster $P$ with size $s$. If (a) chunks $C_1, \ldots, C_v$ are $k^m$-anonymous and (b) the total number of subrecords in all chunks $\sum(|C_i|)$ is greater than or equal to $s + k \cdot (h-1)$, $h = min(m, v)$ or the term chunk is not empty, then Guarantee 1 holds.*

PROOF. To prove this lemma, it suffices to show that for every different combination of $m$ items: (a) no record that contains the $m$ terms can be constructed or (b) a valid initial cluster $P_r$ of size $s$ where the $m$ terms appear in at least $k$ records can be reconstructed.

Assume $m$ random terms $t_1, \ldots, t_m$ from $T^P$. According to Lemma 1, given a disassociated cluster $P_a$, no record that contains these $m$ terms can be created or at least $k$ records can be reconstructed. In the former case, the $k^m$ anonymity trivially holds (this case corresponds to a combination of $m$ terms that did not appear in the initial dataset[1]) and it covers case (a). In the latter case, to prove (b) we need to show that these $k$ records can appear in at least one valid reconstruction of the disassociated cluster $P_a$. A valid reconstruction of cluster $P_a$ is a possible initial cluster that has $s$ non-empty records. We construct a cluster that contains $s$ records in total, where at least $k$ of them contain $t_1, \ldots, t_m$ as follows. We first construct the $k$ records, denoted as $R_k$ that contain the $m$ terms as described in Lemma 1. If the $m$ terms are scattered in $h$ chunks, then to construct each of these records we need $h$ subrecords; one form each chunk, thus $k \cdot h$ subrecords. To create a valid initial dataset of size $s$ that contains the $R_k$ records we only need to populate it with $s - k$ additional records $R_o$ that are valid i.e., non-empty. If the term chunk is non-empty then the $s - k$ records can be populated by randomly combining terms from the term chunk. If the term chunk is empty, we can create such records by assigning 1 subrecord that has not been used in the construction of $R_k$, from any of the $C_1, \ldots, C_v$ chunks,. The total number of subrecords needed is $h \cdot k + s - k = s + k \cdot (h-1)$. The worst case, i.e., the maximum number of subrecords that are required for constructing a valid cluster, is when we need to combine one different subrecord for each of $t_1, \ldots, t_m$ to create a record of $R_k$. In this case, $h = m$ or if the cluster has less than $m$ record chunks $h = v$. Thus, having $s + k \cdot (h - 1)$, $h = min(m, v)$ subrecords is sufficient to create a valid initial cluster. □

---
[1]If a combination of terms cannot be created by combining subrecords, it holds that it did not appear in the original data. The reverse is not true; if a combination can be created, it does not mean that it existed in the original data.

**Joint cluster anonymity.** An example which demonstrates that careless creation of shared chunks can lead to cases where combinations of $m$ terms might not appear $k$ times in any reconstructed dataset is depicted in Figure 5a. Although every chunk (i.e., vertical partition) in the illustrated dataset is $3^2$-anonymous, the overall dataset is not. Since each record has set semantics, an adversary can discard initial datasets that contain records with two identical terms. An attacker $A$ knowing that a user $U$ asked for terms x and o can only find one matching record in every possible original dataset using the following reasoning. Term x appears only in the $1^{st}$ cluster (always together with a) and o appears in the shared chunk. Thus, to construct $U$'s record, $A$ has to combine {a,x} with any of {a,x}, {a} and {o}; but, by the semantics of shared chunks, the only allowed combination is {a,x,o} which appears just once. In order to avoid these conflicts we enforce the following property.

PROPERTY 1. *Let $J$ be a joint cluster and $T^r$ be the set of terms that appear in the record and shared chunks of the clusters (joint or not) forming $J$. A shared chunk of $J$ that does not contain terms from $T^r$ must be $k^m$-anonymous; if it does, it must be $k$-anonymous.*

For example in Figure 5a, Property 1 does not hold since $T^r =$ {a,b,c,d,e,f,x}, term a appears on the shared chunk, a $\in T^r$ and the shared chunk is not $k$-anonymous. On the other hand, the property holds for Figure 5b. $T^r$ contains all terms that appear in $J$ except those that are placed in term chunks and those that appear only in $J$'s shared chunks (only o in the previous example). Let us now consider the following lemma.

LEMMA 3. *A joint cluster for which Property 1 holds, is $k^m$-anonymous.*

PROOF. We will prove the Lemma by induction. Lemma 2 shows that simple clusters are $k^m$-anonymous. It is also easy to see why joint clusters who contain only simple clusters are $k^m$-anonymous, since no conflicts between the terms of the record and shared chunks appear there. In the following we will prove the inductive step; a joint cluster $J$ that is formed by existing $k^m$-anonymous joint clusters is $k^m$-anonymous too.

Let $J$ be a joint cluster with domain $T^J$, the $k^m$-anonymous joint clusters $J_1, \ldots, J_q$ be its children and the simple clusters $k^m$-anonymous $P_1, \ldots, P_w$ be its leaves. Let $\mathcal{SC}$ be the set of the shared chunks of $J$ that all satisfy Property 1. Moreover, let $T^r$ be the set of terms that appear in the record and shared chunks of $J_1, \ldots, J_q$. Since $J_1, \ldots, J_q$ are $k^m$ anonymous we only have to check how the introduction of the shared chunks $\mathcal{SC}$ affects anonymity. Because Lemma 2 holds for each cluster independently, there is no need to set a new bound for the number of subrecords contained in $\mathcal{SC}$. We only have to show that the addition of $\mathcal{SC}$ allows the creation of $k$ records (or no record at all) that contain any $m$-sized combination of terms from $T^J$.

Assume a random combination of $m$ terms $t_1, \ldots, t_m$ from $T^J$ where terms $t_1, \ldots, t_i$ appear in $J_1, \ldots, J_q$ (in either record or term chunks) and $t_{i+1}, \ldots, t_m$ appear in the shared chunks $\mathcal{SC}$. If $i = m$, i.e., all terms belong to $J_1, \ldots, J_q$, then $k^m$ anonymity holds since we assumed that $J_1, \ldots, J_q$ are $k^m$-anonymous. If $i = 0$, i.e., all terms belong to the shared chunks, then by following the same constructive proof as we did in Lemma 1 we can create $k$ records that contain $t_1, \ldots, t_m$. This is sufficient for proving $k^m$-anonymity since there is no requirement for the number of subrecords in the shared chunks. Finally, if some of the $m$ terms cannot appear together by any combination of subrecords, i.e., they did not appear together in the original data at all, then the $k^m$-anonymity trivially holds. It remains to prove that $J$ is $k^m$-anonymous for $0 < i < m$.

Let $SC_1, \ldots, SC_n$, $n \leq m$, with domains $T^1, \ldots, T^n$ be the shared chunks of $\mathcal{SC}$ that contain $t_{i+1}, \ldots, t_m$. Using the reconstructed clusters of $J_1, \ldots, J_q$ we partially reconstruct a joint cluster $J_r$ that contains at least $k$ records with the terms $t_1, \ldots, t_i$. Let $PR$ be the partially reconstructed records of $J$ that contain $t_1, \ldots, t_i$, $|PR| \geq k$. We expand the $PR$ with subrecords from each $SC_i$ of $SC_1, \ldots, SC_n$ to create records that contain all $m$ terms. For each of $SC_i$ with domain $T^i$ we have two cases:

$T^r \cap T^i = \emptyset$ **holds:** In this case, $SC_i$ is $k^m$-anonymous and no term from $SC_i$ appears in any of the $PR$ records. We can then select $k$ subrecords that contain the terms from $t_{i+1}, \ldots, t_m$ that fall in $T^i$ and concatenate them to $k$ records of $PR$.

$T^r \cap T^i \neq \emptyset$ **holds:** In this case, $SC_i$ is $k$-anonymous. Let $SR_i$ be the records of $SC_i$ that contain the terms of $t_1, \ldots, t_m$ that fall in $T^i$, $|SR_i| \geq k$. We want to append $k$ subrecords from $SR_i$ to $k$ records of $PR$ to create records that contain all $m$ terms. Still, not all combinations of $PR \times SR_i$ are valid due to conflicts. The conflicts are caused by terms that appear both in the subrecords of $SC_i$ and the records of $J_r$ that have partially been constructed until now. Assume that the conflict is based on a term $a$. $a$ is independent of $t_1, \ldots, t_m$. Assume that $a$ appears in the record or shared chunks of the simple or joint clusters $\mathcal{J}_a$, which are descendants of $J$. The existence of $a$ in these record chunks means that $SR_a$ did not exist in any of $\mathcal{J}_a$, thus the records of $SR_a$ cannot be combined with any of the records of $\mathcal{J}_a$. Let $JR_a$ be the partially reconstructed records of $\mathcal{J}_a$. Because of the conflict, the adversary knows that if any record of $PR$ belongs to $JR_a$ too, then it cannot be combined with $SR_a$ to create the $k$ records we need. To guarantee $k^m$ anonymity, we must be able to combine at least $k$ records from $PR' = PR \setminus JR_a$ and $SR'_i = SR_i \setminus SR_a$ or none at all. We will prove this by showing that either all records of $PR'$ and all subrecords of $SR'_i$ are disqualified, or that at least $k$ remain in each set. Since each joint cluster is anonymized independently, it contributes at least $k$ records to $PR$. Any conflict with even one record of a cluster from $\mathcal{J}_a$ disqualifies all the records from the same cluster, thus $JR_a$ will be equal to $PR$ or they will differ at least by $k$ records, i.e., all the records contributed by a cluster that has no conflict. Thus $|PR'| = 0 \vee |PR'| \geq k$. Moreover, since we required that $S_i$ is $k$-anonymous, there will be at least $k$ duplicates of each record. A conflict over term $a$ will disqualify at least $k$ records, and if records without $a$ exist in $SR'$ there will be at least $k$ of them. So, after eliminating conflicts, $|SR'| = 0 \vee |SR'| \geq k$. Since both $PR'$ and $SR'$ will have either more than $k$ records or none after eliminating conflicting records, we can either create $k$ records that contain all $t_1, \ldots, t_m$ or no such record. □

The proof is similar for conflicts based on more than one item.

Since a disassociated dataset consists of either joint or simple clusters, Lemmas 2 and 3 are sufficient to prove that the whole dataset is $k^m$-anonymous. We only have to show that the properties required by the previous Lemmas can be guaranteed by the algorithm of Section 4. To guarantee the property required by Lemma 2 we only need to add a check at the end of VERPART that verifies that the cluster contains enough subrecords. If the size limit is not met, then by moving the least frequent item of the record chunks to the term chunk, we guarantee that the conditions set by Lemma 2 are satisfied. This solution is always feasible; at least one term will exist to populate the term chunk in each cluster. To satisfy Lemma 3 the refining algorithm has to check in the creation of a shared chunk, if any of its terms appears in the record chunk of any descendant joint or simple cluster. If this holds, then the chunk must be $k$-anonymous, else it can be $k^m$-anonymous. Since there is always the trivial solution of a record chunk that contains only 1 term, which is both $k$-anonymous and $k^m$-anonymous, the refining algorithm always produces a $k^m$-anonymous dataset.

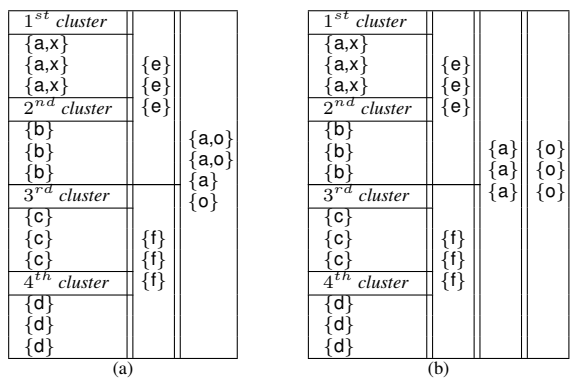

**Figure 5: Unsafe (a) vs. safe (b) creation of a shared chunk**

**Protection against stronger adversaries.** $k^m$-anonymity is a conditional guarantee and the protection it offers is reduced against adversaries with background knowledge that exceeds the attack model assumptions. The most common case is to have adversaries that have more knowledge than $m$ terms about a user or adversaries that have background knowledge about all users that contain certain $m$ terms. In both cases, the adversary's background knowledge consists of enough information to accurately associate some records to a known group of users $\mathcal{U}$. This allows the adversary to remove these records from the groups of candidate records that match her background knowledge for any user who does not belong to $\mathcal{U}$. Still, this attack does not lead automatically to complete re-identification of the additional users, but reduces the number of candidates according to their overlap with the records that are associated with $\mathcal{U}$. This type of attacks has been studied in other contexts [2, 32] and their effect on disassociation and generalization based methods is similar. Disassociation has an additional weakness that is related to Lemma 2; if a record of a user is identified and the remaining terms violate Lemma 2, then the probability of identifying additional records might be reduced to less than $k$-1.

**Diversity.** So far we have discussed an anonymization framework offering protection against identity disclosure. In this section, we discuss how the proposed framework may also offer protection against attribute disclosure and achieve $l$-diversity.

Former works that guarantee $l$-diversity, separate sensitive attributes from quasi identifiers [11, 18, 30]. Following the same idea, we can enforce $l$-diversity in our framework by *(a)* ignoring all sensitive values in the horizontal partitioning and *(b)* placing all sensitive values in the term chunk at the vertical partitioning stage. In the resulting data, all sensitive values will reside at the term chunk and no association between them and any other subrecord or value can be done with probability over than $1/|C|$, where $|C|$ is the size of the cluster. By adjusting the size of the clusters, the anonymization method achieves the desired degree of $l$-diversity.

The proposed anonymization framework offers protection against both identity and attribute disclosure. We focus on the former because to the best of our knowledge there is no other work that employs a similar to disassociation transformation to guarantee protection against identity disclosure (works enforcing $l$-diversity do not consider re-identification dangers [11, 18, 30]). We expect that in practice both protection against identity and attribute disclosure (for the recognized sensitive values) are needed.

## 6. INFORMATION LOSS

By definition, disassociation incurs a different information loss compared to classic anonymization methods. The disassociated dataset preserves all the initial terms and many of the initial itemsets. An analyst can work directly on the disassociated dataset or reconstruct a possible initial one. In the former case, the analyst can compute lower bounds of the supports of all terms and itemsets. These bounds can be computed by counting all the appearances of terms and itemsets in the record chunks of the simple and joint clusters and by adding one to the support of each term that appears in a term chunk. Moreover, the analyst can employ models for answering queries in probabilistic databases to directly query the anonymization result [9]. Using such a model, one can assume that the contents of each record chunk are possible assignments to every record of the cluster with probability $(1/|P|)$. Still, existing work on uncertain data management is not tailored to the disassociated dataset and does not take advantage of the constraints in the reconstruction procedure that we detailed in Section 3 to increase the quality of the estimations. Moreover, working directly on the disassociated dataset requires adjusting existing tools and models for analyzing data. Because of this, we believe that it is easier to apply most analysis tasks on a reconstructed dataset. During horizontal partitioning, clusters are created by bringing similar records together; thus, the statistical properties of a reconstructed dataset are quite close to the original one. A way to further increase the accuracy of the analysis on reconstructed data is to create more than one reconstructed datasets and average the query results on them. We experimentally evaluate the similarity between the reconstructed datasets and the original one in Section 7.

Disassociation hides infrequent term combinations, therefore the incurred information loss is related to term combinations that exist in the original dataset but are lost in the disassociated dataset. To assess the impact of the information loss, we examine the behavior of common mining and querying operations on the transformed data. We employ metrics that are generic and can be used as a comparison basis with anonymization methods that employ different data transformations (such as generalization, suppression and differential privacy). More specifically, we examine how many of the frequent itemsets that exist in the original data are preserved in the published data, and we also measure the relative error in the estimation of the supports of pairs of items.

**Top-$K$ deviation** ($tKd$). The $tKd$ metric measures how the top-$K$ frequent itemsets of the original dataset change in the published anonymized data. Let $FI$ (respectively, $FI'$) be the top-$K$ frequent itemsets in the original dataset (respectively, the anonymized dataset); $tKd$ is defined as follows:

$$tKd = 1 - \frac{|FI \cap FI'|}{|FI|} \qquad (2)$$

Intuitively, $tKd$ expresses the ratio of the top-$K$ frequent itemsets of the original dataset that appear in the top-$K$ frequent itemsets of the anonymized data.

To compare disassociation with generalization-based methods, we define an appropriate version of $tKd$, called the *top-k multiple level mining loss* $tKd$-$ML^2$, which is based on the $ML^2$ metric defined in [27]. Mining a dataset at multiple levels of a generalization hierarchy is an established technique [12], which allows detecting frequent association rules and frequent itemsets that might not appear in the most detailed level of the data. If a generalization hierarchy that allows the anonymization of the data exists, then we can assume that the same hierarchy can be used to mine frequent itemsets from the published (and the original) data at different levels of abstraction. $tKd$-$ML^2$ is given again by Equation 2, but in this case $FI$ and $FI'$ are the sets of generalized frequent itemsets that can be traced in the original and anonymized data, respectively. In the case of generalized datasets, a generalized frequent itemset is lost if it contains terms that have been generalized at a higher level during the anonymization process. Reconstructed datasets do not contain any generalized items, but given a generalization hierarchy generalized frequent itemsets can be mined.

| Dataset | $\|D\|$ | $\|T\|$ | max rec. size | avg rec. size |
|---|---|---|---|---|
| *POS* | 515,597 | 1,657 | 164 | 6.5 |
| *WV*1 | 59,602 | 497 | 267 | 2.5 |
| *WV*2 | 77,512 | 3,340 | 161 | 5.0 |

**Figure 6: Experimental datasets**

**Relative error** ($re$). This metric (used also in [6]) is used to measure the relative error in the support of term combinations in the published data. Since there is a huge number of possible combinations, we limit ourselves to combinations of size two as an indication of the dataset quality. Larger combinations are usually infrequent, and the case of very frequent ones is already covered by $tKd$. The relative error is defined as follows:

$$re = \frac{|s_o(a,b) - s_p(a,b)|}{AVG(s_o(a,b),\ s_p(a,b))}, \quad (3)$$

where $s_o(a,b)$ and $s_p(a,b)$ is the support of the combination of terms $(a,b)$ in the original and in the published data, respectively. Reconstructing anonymized datasets might introduce new item combinations in the published data, which did not exist in the original data. In order to take them into account in the definition of the relative error, we use the average of the two supports as denominator, instead of using the original support $s_o(a,b)$. The average has a smoothing effect on the metric, since it normalizes $re$ to $[0,2]$, and avoids divisions by 0.

## 7. EXPERIMENTAL EVALUATION

The goal of the experimental evaluation is to demonstrate the advantages that disassociation in preserving data quality and to show that disassociation has a robust behavior in different settings.

### 7.1 Experimental Settings

**Datasets.** In the experiments, we use the 3 real datasets described in Figure 6, which were introduced in [33]. Dataset *POS* is a transaction log from an electronics retailer. Datasets *WV*1 and *WV*2 contain click-stream data from two e-commerce web sites, collected over a period of several months. Synthetic datasets were created with IBM's Quest market-basket synthetic data generator (http://www.almaden.ibm.com/cs/quest/syndata.html). Unless otherwise stated, the default characteristics for the synthetic datasets are $1M$ records, $5k$ term domain and 10 average record length.

**Evaluation metrics.** We measure the information loss incurred by our method with respect to the following: *(a)* the $tKd, tKd$-$ML^2$, and $re$ measures defined in Section 6 and *(b)* the percentage of terms $tlost$ that have support more than $k$ in the original dataset $D$ but they are placed in term chunks by our method. We report $tKd$ and $re$ for the disassociated datasets calculated in two different ways: *(a)* one on a single random reconstructed dataset, labeled as $tKd$ and $re$, and *(b)* one calculated only by taking into account the subrecords that appear inside the record and shared chunks, labeled as $tKd$-$a$ and $re$-$a$. In the latter case, we do not take into account the probability that an itemset can be created by combining subrecords. $tKd$-$a$ and $re$-$a$ trace the itemsets that would exist in *any* reconstructed dataset, thus they are based on lower bounds of itemset supports in the original dataset. $tKd$ and $tKd$-$ML^2$ are measured for the 1000 most frequent itemsets. Finally, computing an average $re$ on all combinations of size 2 is not very informative in cases of skewed distributions and large domains. The majority of combinations would be rare or would not exist at all in the original data, but they would dominate the result. To avoid this, we ordered the domain of each dataset by descending term support and we used a small range of consecutive terms to trace their $re$. After some testing we chose the 200th-220th most frequent terms. $re$ in this case is an indicator of how well less frequent but not utterly rare combinations are preserved in the anonymized dataset.

**Evaluation parameters.** We compared performance by varying the following parameters: *(a)* $k$, *(b)* the size of the dataset, *(c)* the size of the dataset's domain, *(d)* the average size of the records, *(e)* the terms we use to calculate $re$ and *(f)* the number of reconstructed datasets we use to calculate $re$ and $tKd$. We do not present a detailed evaluation for $m$, because in all the available datasets its effect for values $m > 2$ is negligible. The explanation for this is that most record clusters are $k^m$-anonymous for any $m$ either because they have gathered very frequent terms or because they contain small subrecords. The experiments are all performed with $k = 5$, $m = 2$ unless explicitly stated otherwise.

**Comparison to state-of-the-art.** Comparing disassociation to other methods is not straightforward; no other method offers the same privacy guarantee while introducing the same type of information loss. We chose to compare disassociation to the generalization-based *Apriori* approach [27], since it offers the same privacy guarantee and it is the most closely related method. This comparison allows us to see how the different data transformations, generalization and disassociation, affect the quality of the anonymized dataset in a similar privacy framework. Furthermore, we compare disassociation to *DiffPart* [6], which offers differential privacy for set-valued data by suppressing infrequent terms and adding noise. The comparison with *DiffPart* demonstrates the gains disassociation offers in terms of information quality, when a more relaxed guarantee like $k^m$-anonymity is chosen over differential privacy. All methods were implemented in C++ (g++ 4.3.2).

### 7.2 Experimental Results

The first experiment (Figures 7a-d) investigates the information loss by our method on the real datasets. In Figure 7a, we see the result of disassociation in the quality of all datasets and in Figures 7b-d just for the *POS* dataset. The $tKd$-$a$ in Figure 7a is similar for all datasets, showing that the most frequent combinations are preserved for different data characteristics. Still, when we trace $tKd$ on the reconstructed datasets, the results significantly improve only for the *POS* dataset, which is the largest of the 3 and its records have the longest average length. This reflects the fact that disassociation managed to create multiple record chunks for *POS*. The combinations of their contents results to a significantly better reconstructed dataset. Disassociation produces significantly different results for the 3 datasets, when looking to the $re$ and $re$-$a$ metrics. The supports of the combinations traced by $re$ are preserved better when the ratio of the dataset size to the dataset domain is high. This ratio is higher for *POS* and *WV1*, where $re$ has significantly superior results to $re$-$a$. This indicates the gains from combining terms from different record chunks in the reconstructed datasets. Finally, the same ratio affects how many terms are placed in the record chunk, as reported by $tlost$, but to a lesser degree. In Figures 7b and 7c, we see how information loss escalates as the power of the guarantee, expressed by the $k$ parameter, grows. The measures that depend on the most frequent items and itemsets are only slightly affected (Figure 7b), since the disassociation algorithm preserves them in record chunks. On the other hand, $re$, which does not depend on the most frequent items, increases linearly with $k$, but with a low rate (Figure 7c). In Figure 7d we explore the gain in information quality by creating several reconstructed datasets and averaging the itemset supports on them. We created 10 random reconstructions of the anonymized *POS* dataset, and we traced $re$ taking this time into account the average supports of the itemsets in 2 ($re$-$r2$), 5 ($re$-$r5$) and 10 ($re$-$r5$) of the reconstructed datasets. We do not report results for $tKd$ since they were already close to 0 and did not benefit substantially from multiple reconstructions. We measure the $re$ on the combinations of the 0-20, 100-120, 200-220, 300-320 and 400-420 most frequent terms in *POS*. In the $x$-axis of Figure 7d, we depict the frequency order of the terms; e.g. a point over 100 refers to the $re$ of the combinations of the 100th-120th most frequent terms in *POS*. When the terms are frequent, the support of their combinations is reported accurately in any reconstructed dataset, so taking the average does not provide any benefit. As the combinations become less frequent, using more than one reconstructed datasets allows for more accurate estimations. In the previous experiments we also examined separately how frequent itemsets of size less or equal to $m$ and of size greater than $m$ ($m = 2$) are preserved. We did not notice any systematic behavior; depending on the dataset, any of the aforementioned frequent itemset classes may be preserved better. For example, frequent itemsets smaller than m were preserved better in *POS* and in *WV2* and worse in *WV1*. We do not report detailed results due to space limitations.

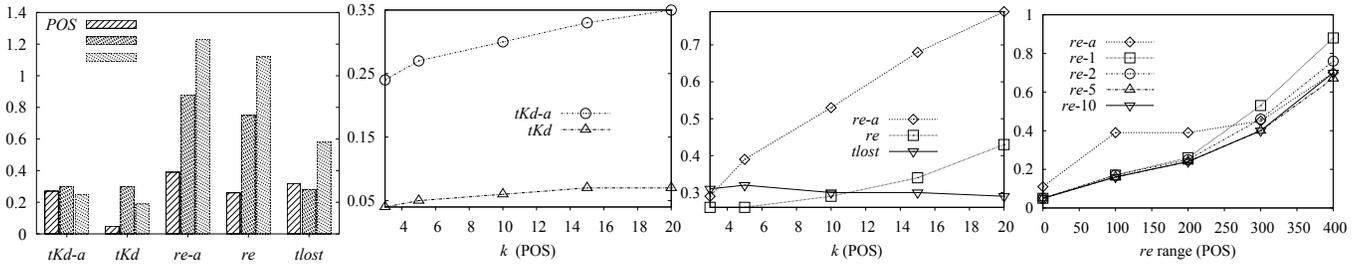

**Figure 7: Information loss on real data (a-d)**

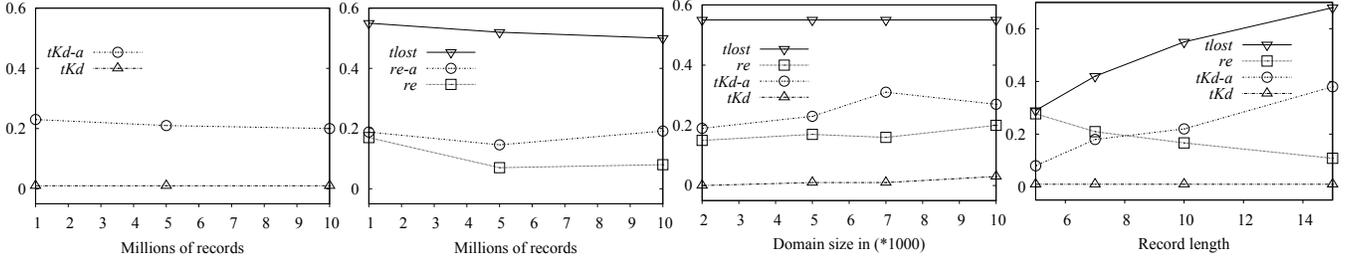

**Figure 8: Information loss on synthetic data (a-d)**

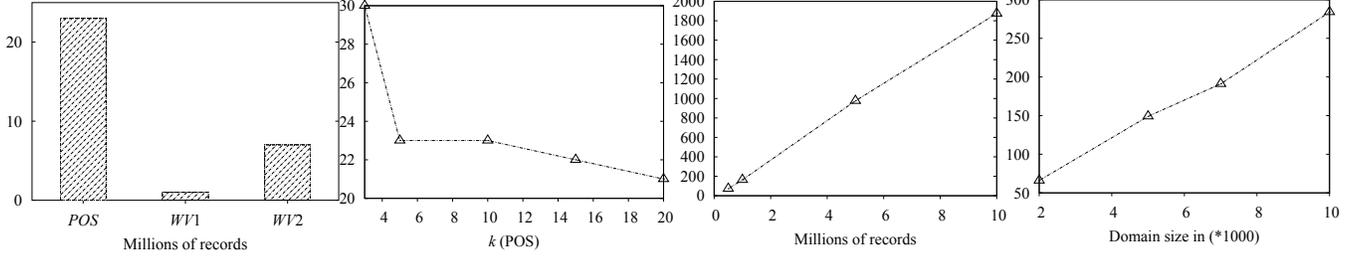

**Figure 9: Performance on real data (a-b)**     **Figure 10: Performance on synthetic data (a-b)**

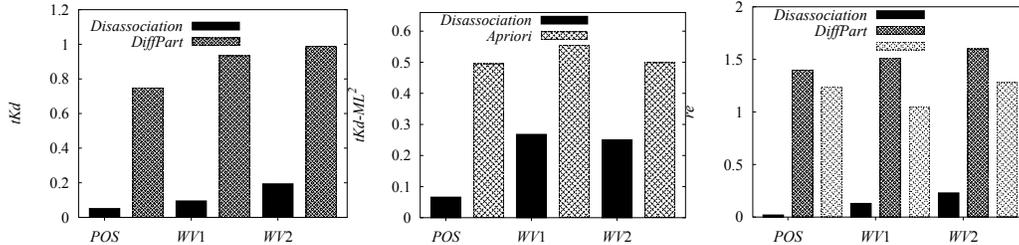

**Figure 11: Comparison with other methods (a-c)**

In the experiments of Figure 8 we used synthetic data to see how the information loss is affected, when the dataset characteristics variate. Since the anonymization is applied independently on each cluster, the database size does not have a significant effect on the quality of the results as demonstrated in Figures 8a and 8b. Only the $re$ and $re$-$a$ are positively affected, because the terms it traces become more frequent and they end up in record chunks more of-ten. Moreover, in Figure 8c we see that increasing the domain when the distribution is skewed, basically affects the tail of the distribution, thus it does not have a significant effect on frequent combinations of terms traced by $tKd$, whereas $re$ slightly deteriorates. The effect of record length is depicted in Figure 8d. Having larger records results in more record chunks and more rare terms in each cluster, thus $tKd$-$a$ and $tlost$ increase. On the other hand, when we keep the dataset and domain size constant and we only increase the record size, the support of the terms in the dataset increases and this explains how $re$ benefits from larger records. Finally, $tKd$ remains close to 0 for all record sizes, since the multiple record chunks reconstruct most of the frequent itemsets in the reconstructed dataset.

Figures 9 and 10 illustrate the performance of the proposed algorithm in terms of CPU time (results in seconds). Disassociation is not significantly affected by the value of $k$, and at the same time it scales linearly to the dataset and the domain size.

Figure 11 shows how disassociation performs compared to *DiffPart* and the *Apriori* algorithm. The graphs illustrate the impact of all algorithms on the quality of the anonymized dataset for $k = 5$ and $m = 2$ (*DiffPart* is unaffected by this parameter). For the *DiffPart* algorithm we used privacy budgets ranging from 0.5 to 1.25, using a step 0.25 with the same parameters as in [6] and we report the best results. In Figure 11a we see how disassociation compares to *DiffPart* in terms of $tKd$. Since in both cases the anonymized datasets contain only original terms (the differential private one has only a subset of them) $tKd$ is computed in exactly the same way. The trade-off for using a stronger privacy guarantee like differential privacy is quite important; in the best case 75% of the top frequent items have been lost, whereas disassociation loses only 5% in the same experiment. In Figure 11b we see how disassociation compares to *Apriori* in terms of $tKd$-$ML^2$, since no original frequent itemset appears in the generalized dataset. Disassociation performs again significantly better than *Apriori* especially for *POS* which is the largest dataset and has more frequent terms than *WV1* and *WV2*. A problem of *Apriori* is that few uncommon terms cause the generalization of several common ones. Finally, Figure 11c shows how all algorithms compare in terms of $re$. $re$ in the generalized dataset is calculated by uniformly dividing the support of a generalized term to the original terms that map to it. *DiffPart* has suppressed all the 200-220th most frequent terms in *POS* (less that 100 of the original 1657 terms are left), so in order to make the comparison meaningful we report the $re$ for the (0-20th) most frequent terms. The $re$ for both *DiffPart* and *Apriori* is over 1, which indicates that the supports of the term combinations have limited usefulness for analysis, whereas disassociation provides 0.18 $re$ in the worst case.

In summary, the experiments on both real and synthetic datasets demonstrate that disassociation offers an anonymized dataset of significantly superior quality compared to other state-of-the-art methods. Moreover, the information loss does not increase aggressively as $k$ increases. Finally, disassociation is not computationally expensive and it is practical for large datasets.

## 8. RELATED WORK

Privacy preservation was first studied in the relational context and focused on protection against identity disclosure. In [25, 26] the authors introduce $k$-anonymity and use generalization and suppression as their two basic tools for anonymizing a dataset. *Incognito* [15] and *Mondrian* [16] are two well known algorithms that guarantee $k$-anonymity for a relation table by transforming the original data using global (full-domain) and local recoding, respectively. [21] demonstrates that the information loss, when providing $k$-anonymity, can be reduced by using *natural* domain generalization hierarchies (as opposed to user-defined ones).

To address the problem of attribute disclosure, where a person can be associated with a sensitive value, the concept of $\ell$-diversity [20] was introduced. *Anatomy* [30] provides $\ell$-diversity and lies closer to our work, in the sense that it does not generalize or suppress the data, but instead it disassociates them by publishing them separately. Still, the anonymization approach is restricted to relational data and it does not protect against identity disclosure. *Slicing*, a more flexible version of *Anatomy* appears in [18]. *Slicing* guarantees $l$-diversity as *Anatomy*, but instead of completely separating sensitive attributes from quasi-identifies, it might publish some quasi-identifiers without disassociating them from sensitive values, if the diversity guarantee is not violated. Moreover, *Slicing*, disassociates quasi-identifiers to increase protection from membership disclosure. By disassociating quasi identifiers, an adversary is faced with several options for reconstructing each record, thus she cannot be certain that a specific record existed in the original data. The data transformation is similar to the approach of our work, but there are significant differences: a) there is no protection against identity disclosure and b) the disassociation between quasi-identifiers does not provide any privacy guarantee, and it takes place only if the impact on information loss is limited. Protection against membership disclosure is facilitated, but not guaranteed; it is roughly estimated using the number of attribute combinations, and not guaranteed by considering the possible initial datasets as in our work. The issues of empty and duplicate records are not addressed. Our work differs from *Slicing* mainly because it uses the disassociation of quasi-identifiers to provide a guarantee against identity, and because it addresses *sparse* multidimensional values.[2].

A similar idea, the vertical fragmentation of relational tables, is employed in a different context to guarantee user anonymity in [7]. The proposed technique distributes a relational table to different servers. In each server, only a subset of the relation's attributes are available unencrypted. The subsets that are available without encryption are chosen so that sensitive associations between attributes, captured by *confidentiality constraints*, are broken. Fragmentation is similar to the basic idea in our work and in [30, 18], but the anonymization model is very different since it focuses on known confidentiality constraints; attacks based on background knowledge are not considered.

More recently, a stronger privacy preservation paradigm, differential privacy, has been proposed [10]. Differential privacy is independent of adversary's background knowledge and it roughly requires that the existence of every single record in the data does not have a significant impact in any query. Finally, the work of [8], although focusing at the protection of associations in sparse bipartite graphs, is related to our work because of the way they define their semantics. The anonymization technique of [8] replaces each node of the graph with a safe group of labels, allowing in this way the anonymized graph to be matched to multiple possible initial graphs.

**Privacy on set-valued data.** The works that lie closer to this paper are those for privacy on set-valued data. Most works that provide protection against identity disclosure rely on generalization. An efficient algorithm for classical $k$-anonymity in a set-value context appears in [13]. [27, 28] introduce the $k^m$ anonymity guarantee, which is used and extended in this paper. The authors provide algorithms for anonymizing the data that, unlike our approach, are based on generalization, employing both local and global recoding. In [4] an algorithm for providing $k^m$-anonymity using only suppression is proposed. The authors have a similar motivation to our work and focus on web search query logs, which they anonymize by removing terms that violate $k^m$-anonymity. The proposed method preserves original terms, but due to the large tail of the term support distribution in such logs, it removes 90% of the terms even for low $k$ and $m$ values. In a different setting, [22] studied multirelational $k$-anonymity, which can be translated to a problem similar to the one studied here, but the anonymization procedure still relies on generalization. [31] provide protection both against identity and attribute disclosure by relying on suppression.

---

[2]In [18] there is an application of *Slicing* to the Netflix data [23], which are sparse. This is achieved by padding all null values with the average of the corresponding attribute values. This technique works only for specific types of data processing and cannot address of sparse data in general.

Protection against attribute disclosure is provided both by generalization and disassociation transformations. The work of [11] extends [30] to provide $\ell$-diversity for transactional datasets with a large number of items per transaction, but it does not depart from the anonymization framework of [30]; it still has a separate set of quasi-identifiers and sensitive values. The basic idea of [11] is to create equivalence classes where the quasi-identifiers are published separately from the sensitive values and their supports. [5] provides a more elaborate $\ell$-diversity guarantee for sparse multidimensional data, termed $\rho$-uncertainty, where sensitive items can act as quasi-identifiers too. Still, unlike our approach, generalization and suppression are employed to anonymize the data.

There have been few works that investigate the publication of set-valued data under differential privacy guarantees. [14] focuses on the anonymization of web search logs, using the AOL data [3]. The proposed method that guarantees differential privacy but it only publishes query terms and not records. Moreover, the anonymization procedure completely hides all terms that are infrequent, which are the majority of terms in AOL data. In [6] a method for publishing itemsets instead of isolated terms from a set-valued collection of data is proposed. The *DiffPart* algorithm follows a top down approach, which starts from the unification of the whole domain and refines it by partitioning it to subdomains, if the item combinations can be published without breaching differential privacy.

Our work lies closer to [11, 30, 18] in the sense that it does not suppress or generalize the data but instead it severs the links between values attributed to the same entity. Unlike [11, 30, 18] we focus on identity protection, and not simply on separating sensitive values from quasi-identifiers. The work of [18] has the most similar data transformation, but it solves a different problem and does not address the peculiarities of sparse multidimensional data. Our privacy guarantee comes from [27], but we follow a completely different path with respect to the data transformation and the type of targeted data utility.

## 9. CONCLUSIONS

In this paper, we proposed a novel anonymization method for sparse multidimensional data. Our method guarantees $k^m$-anonymity, for the published dataset using a novel data transformation called *disassociation*. Instead of eliminating identifying information by not publishing many original terms, either by suppressing or generalizing them, we partition the records so that the existence of certain terms in a record is obscured. This transformation introduces a different type of information loss from existing methods, making it a valuable alternative when the original terms are important.